\documentclass[prx,twocolumn,english,superscriptaddress,floatfix,longbibliography,nofootinbib,preprintnumbers]{revtex4-2}
\usepackage{amsmath,amssymb,amsfonts,graphicx}
\usepackage{ mathrsfs }
\usepackage{dsfont}
\usepackage[caption=false]{subfig}
\usepackage{algorithm}
\usepackage{algpseudocode}
\usepackage{xr-hyper}
\usepackage[bottom]{footmisc}
\usepackage{xcolor}
\usepackage{mathdots}
\usepackage{comment}

\definecolor{darkred}{rgb}{0.801, 0.0, 0.0}
\definecolor{darkgreen}{RGB}{1,50,32}
\usepackage[colorlinks=true, urlcolor=darkred,citecolor=darkred,anchorcolor=darkred,linkcolor=darkred]{hyperref}
\usepackage{ulem}
\def\eqref#1{Eq.$~$\textcolor{darkred}{ (\ref{#1})}}
\def\figref#1{Fig.$\,$\textcolor{darkred}{(\ref{#1})}}
\def\tableref#1{Table$\,$\textcolor{darkred}{(\ref{#1})}}

\usepackage{xcolor}
\usepackage{marginnote}
\usepackage{bbm}
\usepackage{cleveref}

\begin{document}

\preprint{MIT-CTP/5947}

\title{Lattice field theory for superconducting circuits
}

\date{\today }

\def\RLEaffil{Research Laboratory of Electronics, Massachusetts Institute of Technology, Cambridge, MA 02139, USA}
\def\CTPaffil{Center for Theoretical Physics - a Leinweber Institute, Massachusetts Institute of Technology, Cambridge, MA 02139, USA}
\def\AFRLaffil{Air Force Research Laboratory, Information Directorate, Rome, New York 13441, USA}
\def\ARGaffil{Physics Division, Argonne National Laboratory, Lemont, Illinois, 60439, U.S.A}

\author{Joshua Lin}
\affiliation{\CTPaffil}
\affiliation{\ARGaffil}

\author{Max Hays}
\affiliation{\RLEaffil}

\author{Stephen Sorokanich III}
\affiliation{\AFRLaffil}

\author{Julian Bender}
\affiliation{\CTPaffil}

\author{Phiala E. Shanahan}
\affiliation{\CTPaffil}

\author{Neill C. Warrington}
\email{ncwarrin@mit.edu}
\affiliation{\CTPaffil}

\begin{abstract}

Large superconducting quantum circuits have a number of important applications in quantum computing. Accurately predicting the performance of these devices from first principles is challenging, as it requires solving the many-body Schr\"{o}dinger equation. This work introduces a new, general ab-initio method for analyzing large quantum circuits based on lattice field theory, a tool commonly applied in nuclear and particle physics. This method is competitive with state-of-the-art
techniques such as tensor networks, but avoids introducing systematic errors due to truncation of the infinite-dimensional Hilbert space associated with superconducting phases. The approach is applied to fluxonium, a specific many-component superconducting qubit with favorable qualities for quantum computation. A systematic study of the influence of impedance on fluxonium is conducted that parallels previous experimental studies, and ground capacitance effects are explored. The qubit frequency and charge noise dephasing rate are extracted from statistical analyses of charge noise, where thousands of instantiations of charge disorder in the Josephson junction array of a fixed fluxonium qubit are explicitly averaged over at the microscopic level. This is difficult to achieve with any other existing method.

\end{abstract}
\maketitle

\section{Introduction }

Superconducting electrical circuits with many components form a number of important quantum technologies, including qubits \cite{PhysRevA.76.042319,Manucharyan_2009}, analog quantum simulators \cite{Rosen_2024,rosen2024_flatband}, quantum sensors \cite{PhysRevX.14.011007,danilin2021_quantumsensing,PhysRevLett.131.211001}, amplifiers \cite{9134828,Roy_2016,Esposito_2021,Macklin_science2015}, and transducers \cite{LaHaye:2009psh, Gustafsson_2014,osti_10231749}. Designing, operating, and improving these devices requires a precise theoretical and quantitative understanding of their properties~\cite{fluxonium_tunable_coupler, PRXQuantum.5.020326,PhysRevX.13.031035,Berke_2022}. As superconducting circuits continue to rapidly increase in complexity, characterization is becoming both central and challenging.

This work presents a general method for solving the many-body Schr\"odinger equation arising from circuit-QED, the \textit{de facto} microscopic theory of superconducting circuits.  The approach consists of a direct adaptation of lattice quantum field theory (lattice-QFT) methods to this setting, and arises from viewing a quantum electrical circuit as an instantiation  of a lattice field theory. Quantum dynamics is formulated in the langauge of path-integrals, and the many-body spectrum is extracted from Euclidean-time correlation functions, as typically applied in nuclear and particle physics. The method can be applied to any superconducting circuit with a finite number of capacitors, inductors and Josephson junctions, and is therefore widely applicable. 
While there is a long history of the use of path integrals and Monte Carlo methods in quantum electrical circuits dating back to the 1960s~\cite{PismaZhETF.3.141,PhysRevB.3.762,PhysRevB.6.1767,PhysRevB.10.4598,PhysRevB.15.2828}, all previous work has focused on understanding the bulk thermodynamic behavior of Josephson junction (JJ) arrays. This includes extensive studies of the superconductor/insulator/normal metal phase transitions of both 1d and 2d JJ arrays \cite{SIMANEK1979419,osti_7061039,PhysRevB.30.1138,PhysRevB.49.12115,PhysRevB.54.12361,PhysRevB.101.024518}, dissipative phase transitions \cite{PhysRevLett.94.157001,PhysRevLett.56.2303}, the effects of disorder \cite{PhysRevB.41.8749,PhysRevB.37.5966,ALSAIDI2004216} and frustration \cite{PhysRevLett.51.1999,VANOTTERLO1994504,PhysRevB.54.10081,PhysRevB.65.172505}. 
In contrast, the method developed here allows the computation of not only critical phenomena, but also the entire spectrum of the theory. From the spectrum a wealth of information of practical importance in quantum computing is obtained. This includes the properties of individual, isolated qubits, but also that of qubits integrated into large quantum processors. Matrix elements allow for the prediction of coherence times and cross talk. 

The capabilities of the lattice method developed here are demonstrated via an application to fluxonium, a superconducting qubit which is itself a many-body system. Fluxonium is a promising candidate for scalable quantum computation due to its long coherence times and fast, high-fidelity gates \cite{Nguyen_2019,PhysRevLett.130.267001,PhysRevX.13.031035,PRXQuantum.5.020326}. In a fluxonium-based processor, or even a single fluxonium qubit, computing relevant observables from first principles is prohibitively expensive via direct diagonalization because the circuit typically contains a hundred or more Josephson junctions, each hosting a $U(1)$ quantum variable with an unbounded local Hilbert space. In other words, a complete description of fluxonium requires treating the phase of each island in the qubit as a separate (periodic) quantum degree of freedom.

In practice, fluxonium is commonly studied by substituting the full theory for an approximate, single-variable theory called the \textit{superinductance model}, which is more easily solvable \cite{Manucharyan_2009, symmetries-and-collective}. Indeed, most fluxonium qubits are fabricated precisely so that  the superinductance model describes the device well. However, this restricts exploration of fabrication space and moreover is sometimes insufficient to describe existing fluxonium qubits. For example, as shown in recent experiments, dephasing rates increase dramatically at high impedance, even at fixed qubit parameters \cite{randeria2024dephasing}. This dephasing is understood to arise from coherent quantum phase slips (CQPS), a many-body effect that goes beyond the superinductance model \cite{PhysRevB.85.024521, randeria2024dephasing}. Ground capacitances also produce parasitic interactions that affect the qubit spectrum. 

Microscopic methods applied to investigations of fluxonium include array mode perturbation theory \cite{PhysRevX.3.011003,Viola_2015,Sorokanich:2024wkx},  periodic gaussian states \cite{PhysRevB.102.014512}, and tensor networks (TN) \cite{PhysRevB.102.014512,PhysRevB.100.224507,Di_Paolo_2021}. The lattice technique introduced here is a new addition to this landscape. It is general, and retains its validity at large coupling, which places it at an advantage over perturbative approaches. While most comparable in computational efficiency and precision to the TN approach, the lattice method offers a distinct advantage: it naturally accommodates the full U(1) Hilbert spaces associated with each superconducting phase variable in the circuit. This is achieved by performing Monte Carlo sampling directly over the compact U(1) field. In contrast, TN methods are formulated in terms of qudits, requiring a choice of basis and truncation of the local Hilbert space. As a result, accurate TN simulations rely on a good physical understanding of which basis states are most relevant, to avoid introducing systematic error. Moreover, after an initial lattice calculation of a specific device, the ``reweighting" technique allows one to simulate nearby devices in fabrication space at marginal computational cost. It can be expected that this specific feature will be useful in future prototyping and design of quantum hardware.

To demonstrate the applicability of the lattice approach presented here, three distinct studies are performed. First, the results of a previous TN calculation of fluxonium at high impedance \cite{Di_Paolo_2021} are reproduced. This is the first independent verification of the TN studies of fluxonium. Then, two phenomenological studies are undertaken which are intended to mimic the experimental work of Ref.~\cite{randeria2024dephasing}. Anchored to a set of device parameters explored in this experiment, impedance and ground capacitance in fluxonium are independently tuned, and their effects on fluxonium are explored. Emphasis is placed on exploring many-body effects on the qubit frequency and its associated charge dispersion, which is generated by coherent quantum phase slips interfering with local gate charges via the Aharonhov-Casher effect, and causes dephasing  \cite{Matveev_2002, manucharyan2012superinductance,PhysRevB.85.024521,randeria2024dephasing}. Deviations from current analytic predictions for the dephasing rate are seen, and the lattice results are used to formulate a modified prediction which accounts for fluxonium's distinguished small junction. At very high impedances, multiple phase slip events beyond the scope of current analytic theories appear. The lattice study comprises a statistical analysis of thousands of fluxonium qubits, all simulated according to the microscopic theory, which is likely not obtainable by any other microscopic approach. The behavior of fluxonium as a function of ground capacitance corroborates results from array mode perturbation theory, an analytic approach to computing many-body effects in fluxonium. Interestingly, it is seen that ground capacitances reduce charge noise. A thorough understanding of ground capacitance effects is useful not only in fluxonium, where it causes parasitic interactions with ``array modes", but also in the design of next-generation qubits that rely on long JJ arrays such as the $0-\pi$, blochnium, bifluxon, gridium, and harmonium qubits~\cite{PRXQuantum.2.010339,Pechenezhskiy_2020,PRXQuantum.1.010307, nguyen2025superconducting, hays2025non}. This calculation goes beyond the state-of-the-art achieved with TN, which has yet to include ground capacitances in simulation (though there appears to be no fundamental obstruction to doing so).

The remainder of this manuscript proceeds as follows. Sec. \ref{sec:theory} develops a lattice field theory approach for superconducting circuits. Sec. \ref{sec:fluxonium} discusses the microscopic theory of fluxonium, as well as relevant phenomenology for the circuit. Sec. \ref{sec:results} presents the results of the studies performed in this work. Sec. \ref{discussion} offers further discussions and Sec. \ref{sec:conclusion} concludes. SI units are employed throughout.

\section{Theory}
\label{sec:theory}

This section presents a new method for computing the properties of general superconducting quantum circuits.  It involves formulating the dynamics of the system with circuit QED, expressing the resulting many-body system with path integrals, then computing these path integrals numerically with Monte Carlo methods. Special emphasis is placed on \textit{Euclidean-time correlation functions}, which contain the spectrum of the many body theory. This approach reduces both memory and computational demands compared to exact diagonalization and has systematically-improvable uncertainties.

\subsection{Circuit quantization}
\label{sec:circuit-quantization}

Consider an arbitrary quantum electrical circuit composed of a finite number of capacitors, Josephson junctions, and inductors. A capacitive, inductive, or Josephson element connecting nodes $x$ and $x'$ respectively contribute a term to the circuit's classical Lagrangian proportional to
\begin{equation}
    \frac{C}{2} (\dot{\Phi}_x - \dot{\Phi}_{x'})^2, \frac{1}{2L} ({\Phi}_x - {\Phi}_{x'})^2, -E_J \text{cos}\big(\frac{\Phi_x - \Phi_{x'} }{\Phi_0}\big)~,
\end{equation}
where $\Phi_x$ is the generalized flux on node $x$, $\dot{\Phi}_x$ is its time derivative, and $C,L,E_J$ are respectively the capacitance, inductance, and Josephson energy of the given branch element. Using the method of nodes, all redundant circuit variables are eliminated and a Lagrangian for the circuit is computed, which has the following general form:
\begin{equation}
    L = \frac{1}{2} \sum_{x x'} \dot{\Phi}_x C_{x x'} \dot{\Phi}_{x'} - U(\Phi)~,
\end{equation}
where $C$ is the capacitance matrix of the circuit and $U$ is the total potential energy  (for reviews on the topic see e.g. Refs.~\cite{Vool_2017,quantum_engineers_guide,blais_rmp2021}). The circuit Hamiltonian is constructed by first defining canonically conjugate momenta
\begin{equation}
    Q_x = \frac{\partial L}{\partial \dot{\Phi}_x}\, ,
\end{equation}
then performing a Legendre transformation. The resulting Hamiltonian has the general form
\begin{equation}
    H = \frac{1}{2}\sum_{x x'} Q_x C^{-1}_{x x'} Q_{x'} + U(\Phi)~,
\end{equation}
where $C^{-1}$ is the inverse capacitance matrix. The conjugate momentum $Q_x$ has the physical interpretation of the amount of excess charge on node $x$ relative to a reference configuration in the infinite past. An important modification to the Hamiltonian occurs in the presence of ``gate charges", which are produced by random voltages on nodes. In this case the Hamiltonian is modified to 
\begin{equation}\label{eq:Hflux}
    H = \frac{1}{2}\sum_{x x'} (Q_x - Q_{gx})C^{-1}_{x x'} (Q_{x'} - Q_{gx'}) + U(\Phi)~,
\end{equation}
where $Q_{gx}$ are real-number valued gate charges. The lattice formalism presented in the following subsection is general enough to handle this case.

This theory is quantized by promoting $\Phi_x$ and $Q_x$ to operators satisfying canonical commutation relations $[\Phi_x, Q_{x'}] = i \hbar \delta_{x x'}$. After quantization, $\Phi_x$ and $Q_x$ can be identified as lattice quantum fields, with the collections $\{\Phi_x\}$ and $\{Q_x\}$ defined as ``field operators". These are understood to be associated with nodes of the circuit and generate a many-body Hilbert space. It is convenient to define dimensionless fields
\begin{align}
    \phi_x & = \frac{\Phi_x}{\varphi_0} \\
    q_x & = \frac{Q_x}{2e}~,
\end{align}
which satisfy $[\phi_x,q_{x'}]=i\delta_{x x'}$.
Here $\varphi_0 = \hbar/(2e)$ is the reduced flux quantum and $2e$ is the electric charge of a Cooper-pair. The variables $\phi_x$ and $q_x$ are respectively equal to the phase of the superconducting wavefunction and excess number of Cooper pairs on node $x$.

In the Schr\"odinger picture, the $\phi$-space wavefunction has as many arguments as there are nodes $\psi = \psi(\phi_1,\dots,\phi_N)$. This feature produces an exponential scaling in both memory and computational time for numerical diagonalization techniques. To evade this exponential scaling, the lattice approach developed here trades the Schr\"odinger picture for a Euclidean path integral formulation which is then solved with Monte Carlo methods. This results in polynomial storage requirements and (very often) polynomial computation time. These computational improvements are the main motivations for this method. It is important to note that lattice methods are best suited for  \textit{static} observables, such as energy levels and matrix elements; real-time problems often require exponential computation time to solve with lattice methods because of the ``sign problem" \cite{PhysRevLett.94.170201}.

\subsection{Lattice field theory formulation}
\label{sec:lattice-formulation}

The lattice field theory approach to quantum circuits (and many-body quantum systems in general) relies on formulating the theory in the language of path integrals and extracting physical observables from these. A fundamental observation is the following: the spectrum of the theory can be extracted from fits to Euclidean-time separated correlation functions. Two-point Euclidean-time correlation functions are defined as
\begin{equation}
\label{eq:corr-fcn}
    \langle \mathcal{O}(t) \mathcal{O}(t') \rangle = \frac{1}{\mathcal{Z}} \text{tr} \Big( ~e^{-\beta H} \mathcal{O}(t) \mathcal{O}(t') \Big)~,
\end{equation}
where $\beta^{-1} = k_B T$ is the fictitious inverse temperature of the system, $\mathcal{Z} = \text{tr} e^{-\beta H}$ is the partition function, $\mathcal{O}(t)$ are operators formed from functions of the microscopic fields $\phi_x$ and $q_x$, and where the operators are considered in the Heisenberg picture:
\begin{equation}
\label{eq:time-ev}
    \mathcal{O}(t) = e^{+\frac{H t}{\hbar} } \mathcal{O}(0) e^{- \frac{H t}{\hbar} }~.
\end{equation}
This definition is typically called \textit{Euclidean} time evolution; it does not correspond to physical time but is simply a computational tool. In the energy eigenbasis, the two-point Euclidean-time correlation functions can be expanded as
\begin{align}
\label{eq:twopointspec}
    \langle \mathcal{O}(t) \mathcal{O}(t') \rangle = \sum_{nm}e^{-\beta E_n} & e^{-(E_m-E_n)(t-t')/ \hbar } \times \nonumber \\
    & \langle n | \mathcal{O} | m \rangle  \langle m | \mathcal{O} | n \rangle ~.
\end{align}
The right-hand side shows that if correlation functions can be precisely computed (typically numerically), many-body energies and matrix elements can be extracted.

A path integral representation of correlation functions can be obtained by Trotterizing the Euclidean time evolution into discrete time steps, separating the kinetic and potential energy terms using the Baker-Campbell-Hausdorff (BCH) formula, inserting many complete sets of field- and momentum-space eigenstates $\{|\phi_x\rangle, |q_x \rangle \}$, then explicitly performing the momentum integrals. What remains is a representation of correlation functions expressed as a ``sum over all paths" in field space. The details of this calculation are straightforward and may be found, e.g., in Ref.~\cite{PhysRevB.54.12361}. Key steps are  elaborated below.

The Trotterization is achieved in two steps. First, $e^{-\beta H}$ is trivially rewritten in the following way:
\begin{equation}
    e^{-\beta H} = \prod_{t=1}^{N_t} e^{-\frac{H \Delta t}{\hbar}}  ~,
\end{equation}
where $N_t$ is the number of ``time slices" and $\Delta t = \hbar \beta/N_t$ is the ``temporal lattice spacing", or simply the lattice spacing, which has units of time. When the number of timeslices grows at fixed $\beta$, $e^{-\frac{H \Delta t}{\hbar}}$ approaches to the identity and the following BCH decomposition is a good approximation:
\begin{equation}
\label{eq:BCH}
    e^{-\frac{U \Delta t}{2\hbar}} e^{-\frac{T \Delta t}{\hbar}} e^{-\frac{U \Delta t}{2\hbar}} = e^{-\frac{H_\mathrm{eff} \Delta t}{\hbar}} \approx e^{-\frac{H \Delta t}{\hbar}}.
\end{equation}
Here $T,U$ are respectively the kinetic and potential energy of the circuit under consideration, and 
\begin{equation}\label{eq:Hdt}
    H_\mathrm{eff} = H + O(\Delta t^2)
\end{equation}
is the transfer matrix. 
\Cref{eq:Hdt} implies that the spectrum extracted from lattice calculations have $O(\Delta t^2)$ errors, which can be fit and removed with computations at multiple values of $\Delta t$.

As discussed above, the spectrum is extracted from Euclidean-time correlation functions, which in the lattice theory take the form
\begin{align}
\label{eq:fund-expression}
    \langle \mathcal{O}_x(t) \mathcal{O}_{x'}(t') \rangle_{\text{lat}} = \frac{\int D \phi ~ e^{-\frac{S(\phi)}{\hbar} }   \mathcal{O}_x(t) \mathcal{O}_{x'}(t')}{\int D \phi ~ e^{-\frac{S(\phi)}{\hbar} }}~,
\end{align}
where $S$ is the ``Euclidean action". For the Hamiltonian \Cref{eq:Hflux}, $S$ is given by 
\begin{align}
\label{eq:lat-act}
    S & =  \frac{1}{2} \varphi_0^2 \sum_{t, x, x'} \Delta t  ~\dot{\phi}_{x,t}C_{xx'}\dot{\phi}_{x,t} + \sum_{t,x} \Delta t ~U(\phi_{x,t}) \nonumber \\ 
    & +i \hbar \sum_{t,x} \Delta t ~ q_{gx} \dot{\phi}_{xt} ~.
\end{align}
Here $\dot{\phi}_{x,t} = (\phi_{x,t+1}-\phi_{x,t})/\Delta t$ is a discrete time derivative, $q_{gx} = Q_{gx}/(2e)$ is the gate charge in units of the Cooper pair charge,  and $D\phi = \prod_{t=1}^{N_t}\prod_{x=1}^N d\phi_{xt}$ is a shorthand for the product of all field variables on a discrete spacetime lattice generated by the $N_t$ timeslices of the circuit.

Note that \eqref{eq:lat-act} is the Euclidean action for \textit{any} with Hamiltonian of the form of \Cref{eq:Hflux}. All that is required to employ the lattice formalism is knowledge of the capacitance matrix, the potential energy, and the compact/non-compactness of the circuit variables.

The lattice theory always has discretization errors, however these can be extrapolated to zero in a controlled way by performing calculations at multiple values of $\Delta t$ and fitting results to known scaling laws, such as \Cref{eq:Hdt}. This \textit{a priori} knowledge of the functional form of the dominant systematic error, and leveraging that knowledge to extrapolate this error to zero, is a distinguishing feature of the lattice approach over TN and exact-diagonalization \color{black} in a truncated Hilbert space \color{black}, and is one of the main theoretical advantages of the lattice QFT approach. 

The lattice correlation function in \eqref{eq:fund-expression} may be evaluated using Monte Carlo methods, which are based on the observation that correlation functions can be viewed as averaging against a probability distribution
\begin{equation}
    p(\phi) = \frac{e^{-\frac{S(\phi)}{\hbar} }}{\int D\phi ~e^{-\frac{S(\phi)}{\hbar}} }~.
\end{equation}
The approach taken here is standard in nuclear and particle physics, and can be found in any standard textbook on lattice field theory \cite{Montvay:1994cy,Gattringer:2010zz}. However, since here lattice field theory is applied in a non-standard context, some of the basics are reviewed.

The required probability distribution is sampled by generating a Markov chain in field space that satisfies reversibility, ergodicity, and detailed balance. This guarantees convergence of the Markov Chain to the desired probability distribution \cite{Montvay:1994cy,Gattringer:2010zz}. Correlation functions (such as the following two-point example) are estimated as
\begin{equation}
\langle \mathcal{O}_x(t) \mathcal{O}_{x'}(t') \rangle_{\text{lat}} \simeq \frac{1}{N_{\text{cfg}}}\sum_{i=1}^{N_{\text{cfg}}}\mathcal{O}([\phi_i]_{x,t}) \mathcal{O}([\phi_i]_{x',t'})
\end{equation}
where $\phi_i$ is the $i^{th}$ field along the Markov chain. The numerical estimates of observables converge to the mean at a rate of $N_{\text{cfg}}^{-1/2}$. 

The following strategy is adopted. At any given lattice spacing, statistics are increased until the statistical uncertainties are less than, e.g., a few percent. This is done for at least three different lattice spacings, and the resulting observables are fit to extrapolate to the continuum limit. The final result has both statistical and systematic errors, arising respectively from the finite sampling and lattice spacing extrapolation, and a crucial aspect of the lattice method is that the latter are systematically improvable by performing calculations at finer and finer lattice spacing.

The Monte Carlo method must be modified when gate charges are present, because in this case $e^{-S}$ is complex-valued and is therefore not a probability distribution. In this case observables can be computed using ``reweighting", which amounts to the following observation:
\begin{align}
\label{eq:reweighting}
    \frac{\int e^{-(\text{Re}S + i \text{Im}S)} \mathcal{O}}{\int e^{-(\text{Re}S + i \text{Im}S)} } 
    & = 
    \frac{\int e^{-\text{Re}S } \Big( e^{-i \text{Im}S}\mathcal{O}\Big) / \int e^{-\text{Re}S}}{\int e^{-\text{Re}S} \Big( e^{-i \text{Im}S} \Big) / \int e^{-\text{Re}S} } \nonumber \\
    & = \frac{\langle e^{-i \text{Im}S}\mathcal{O}\rangle_{\text{ReS}}}{\langle e^{-i \text{Im}S} \rangle_{\text{Re}S}}~,
\end{align}
where $\langle \cdot \rangle_{\text{Re}S}$ denotes averaging against the probability distribution $p\sim e^{-\text{Re}S}$. Thus the effect of gate charges, which is to produce an oscillating phase, can always be absorbed by ``reweighting" the observable. 
In practice the strategy for computing observables in the presence of gate charges is to perform a Monte Carlo calculation that samples $p\sim e^{-\text{Re}S}$, specifying a set of gate charges, then computing the final line of \eqref{eq:reweighting}. Observables with \textit{any} set of gate charges can be computed from this initial Markov chain, and this is leveraged in Sec. \ref{sec:study2} to perform an explicit charge disorder averaging with a single calculation. 

Reweighting can be used to incorporate \textit{any} change in the action without having to perform a new calculation, not just to incorporate gate charges. All that is required is to incorporate the corresponding exponential factor into observables. This may be useful in iterative device design, where an initial Monte Carlo simulation may serve as a baseline for subsequent iterations. While always valid in principle, reweighting ceases to be practically useful if the difference between probability distributions between devices is too significant.

\section{Fluxonium}
\label{sec:fluxonium}

\begin{figure}[t!]
    \centering
    \includegraphics[width=0.3\columnwidth]{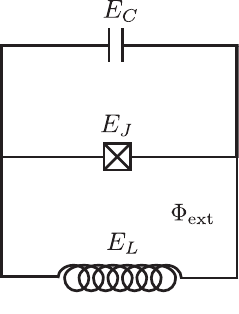}
    \includegraphics[width=0.525\columnwidth]{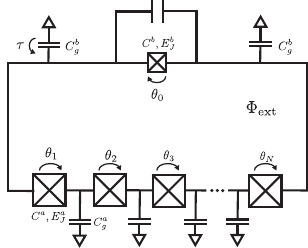}  
    \caption{\textbf{\textit{Fluxonium}}: \textit{(left)} Lumped-element circuit model of fluxonium  
    \textit{(right)} Microscopic circuit model of fluxonium. \textit{Description:}  The microscopic model of fluxonium considered here involves phase differences across junctions $\theta_i$, ground capacitances, and identical array junctions. The lumped-element model approxiates the Josephson junction array as a linear inductor.
    \textit{Notation:} The variables $C^b,C^b_g,E_J^b$ are the capacitance, ground capacitance and Josephson energy of the small junction, while $C^a,C^a_g,E_J^a$ are the same for the array junctions. The model studied includes $N$ array junctions and an external flux $\Phi_{\text{ext}}$ threading the loop. }
    \label{fig:fluxonium}
\end{figure}

As briefly discussed in the introduction, fluxonium is a superconducting qubit which has exhibited favorable properties for quantum computing including high coherence times and fast, high-fidelity gates. It consists of a Josephson junction shunted by a superinductor and a capacitor. Illustrated in the right panel of \figref{fig:fluxonium}, the fine-grained model of fluxonium considered here consists of a superinductor formed by  a one-dimensional array of identical Josephson junctions\footnote{There are other ways to form a superinductance, including granular aluminum and disordered nanowires \cite{Rieger_2022,PhysRevLett.122.010504}. }. Every Josephson junction in fluxonium hosts a $U(1)$ quantum variable equal to the phase difference across it, and fluxonium qubits often have a hundred or more Josephson junctions. It is thus a many-body system. 

The circuit parameters of this device are the capacitance, ground capacitance, and Josephson energy of all junctions: for the small junction these are denoted $C^b,C^b_g, E^b_J$, and for the array junctions by $C^a,C^a_g,E^a_J$. 
While here all array junctions are taken to be identical, it would be straightforward to include disorder if desired. 
This study considers $N$ array junctions and an external flux $\Phi_{\text{ext}}$ threading the loop. Any additional shunt capacitance for the small junction is assumed to be included in $C^b$.

Fluxonium is often approximated by the lumped element model shown in the left panel of \figref{fig:fluxonium}, where the Josephson junction array is approximated by a linear inductor \cite{Manucharyan_2009}. The Hamiltonian of the lumped-element system is 
\begin{equation}
\label{eq:superinductance-model}
    H = 4 E_C n^2 + \frac{E_L}{2} \varphi^2 - E_J \text{cos}(\varphi - \varphi_{\text{ext}})~.
\end{equation}
In this model, the quantum dynamics is approximated by a single quantum variable, the phase drop $\varphi$ across the  small junction alone, and the ``coarse grained" parameters $E_J,E_C,E_L$ emerge from the fine-grained theory. It is common to relate these to microscopic parameters through the relations $E_J = E_J^b$, $E_C = e^2 / (2 C^b)$, $E_L = E_J^a/N$. These relations are sufficient to describe most fabricated devices, however they are approximate and can be modified by many-body interactions. Many-body interactions can also decohere the qubit \cite{randeria2024dephasing}.

This work goes beyond the single-variable model illustrated in the left panel of  \figref{fig:fluxonium} in favor of a more complete, ``microscopic'' theory.  This theory is then used to probe  \textit{coherent quantum phase slips} and \textit{array modes}, two many-body effects in fluxonium. The microscopic theory corresponding to the right panel of \figref{fig:fluxonium} can be deduced from circuit quantization, discussed in Sec. \ref{sec:circuit-quantization}. Here the theory is formulated in terms of ``branch variables", which are denoted $\theta_x,~n_{x}$, rather than node variables $\phi_x,~q_{x}$,  discussed previously. This is simply for convenience and changes nothing about the structure of the formalism. The position-like variable $\theta_x$ is the phase drop across junction $x$, and its conjugate momentum $n_x$ is the number of cooper pairs that have tunneled across the junction. These satisfy $[\theta_x,n_{x'}]=i \delta_{x x'}$ . 

The Hamiltonian of the circuit is
\begin{align}
\label{eq:flux-ham}
    H & = 2 e^2 \sum_{x x'} (n_x - n_{gx}) C^{-1}_{x x'} (n_{x'} - n_{g x'})  - E_J^a \sum_{x} \text{cos}\theta_x \nonumber \\
    &- E_J^b \text{cos}\Big( \sum_x \theta_x - \varphi_{\text{ext}} \Big)~,
\end{align}
where $x,x' = 1,\dots,N$ label the array junctions, $C$ is the capacitance matrix of the circuit, and $n_{gx}$ are local gate charges expressed in the branch variable basis. In defining the Hamiltonian, the fluxoid quantization condition $\theta_0 + \sum_x \theta_x + \varphi_{ext} = 2\pi n$ has been imposed, which eliminates the small junction variable $\theta_0$ in favor of the total phase drop across the array. With these choices only the $N$ array variables enter the theory. The capacitance matrix can be written in the following way:
\begin{equation}
\label{eq:cap-mat-form}
    C = C_1 + C_2 + C_3 + C_4~.
\end{equation}
Here $(C_1)_{xx'} = C^a \delta_{x x'}$, $(C_2)_{xx'} = C^b + C^b_g$, 
\begin{equation}
\label{eq:C3}
    C_3 = C^a_g 
    \begin{bmatrix}
    N-1&N-2&\dots&1&0\\
    N-2&N-2&\dots&1&0 \\
    \vdots&\vdots&\ddots&\vdots\\
    1&1&\dots&1&0\\
    0&0&\dots&0&0\\ 
    \end{bmatrix},
\end{equation}
and $C_4 = - \vec{b} \vec{b}^T/a$, with $a = 2 C^b_g + (N-1) C^a_g$ and $(\vec{b})_x = C^b_g +C^a_g(N-x)$ \cite{Sorokanich:2024wkx,Viola_2015,symmetries-and-collective,Di_Paolo_2021}. Note that the chosen capacitance matrix \eqref{eq:cap-mat-form} encodes a floating fluxonium; a grounded device is obtained by setting $C = C_1 + C_2 + C_3$.

\section{Results}
\label{sec:results}

Three distinct numerical studies of fluxonium are undertaken, spanning validation of the lattice method by comparison with established results for a specific system, through new explorations of fluxonium's parameter space. First, a device previously studied with TN \cite{Di_Paolo_2021} is investigated using the lattice approach and precise agreement is seen. In addition to validating the lattice method, this study constitutes the first independent check of Ref.~\cite{Di_Paolo_2021}. Second, theoretical calculations are undertaken relating to the  experimental work of Ref.~\cite{randeria2024dephasing}, where a suite of fluxonium qubits with varying array junction impedances were fabricated with (nearly) fixed $E_C,E_J,E_L$. Theoretical calculations are performed using the circuit parameters of the highest impedance device reported in \cite{randeria2024dephasing};
the qubit frequency, charge dispersion, and coherent quantum phase slips are explored. Anchored on this device, two independent directions in parameter space are explored, in particular higher impedance and non-zero ground capacitance. A quantitative analysis at non-zero ground capacitance has not been accomplished with any many-body method to date, and many-body effects not accounted for in the superinductance model are seen.

\subsection{Study 1: Validation}
\label{sec:study1}

This study considers a device previously analyzed via TN in Ref.~\cite{Di_Paolo_2021}, with  microscopic circuit parameters: 
\begin{align}
\label{eq:tn-comp-params}
    N &= 40 \nonumber \\
    C^b & = 7.5 \text{ fF} \nonumber\\
    E^b_J & = 8.9 \text{ h GHz} \nonumber \\ 
    \hbar \omega_{\text{pl}} & = 12.5 \text{ h GHz} \nonumber \\
    z & = 0.14~.
\end{align}
Here $z = \pi^{-1}\sqrt{2 E^a_C / E^a_J}$ and $\hbar \omega_{\text{pl}} = \sqrt{8 E^a_C E^a_J}$ are respectively the \textit{reduced impedance} and \textit{plasma frequency} of the array junctions. The external flux is fixed to $    \varphi_{\text{ext}} = \pi $.

The observable of primary interest is the qubit frequency $\nu_{01} = (E_1-E_0)/h$ and its dependence on the local gate charges. Here $E_0,\, E_1$ are the ground and first-excited states of the full many body theory. The qubit frequency is extracted from the lattice theory through the correlation function
\begin{equation}
\label{eq:eff-freq}
    C(t) = \langle \mathcal{O}(t) \mathcal{O}(0)\rangle ~,
\end{equation}
where $\mathcal{O} = \sum_{x=1}^N \text{sin }\theta_x$ is an ``interpolating operator" with good overlap with the $0\rightarrow 1$ transition of the many body theory. That $\mathcal{O}$ is a reasonable operator for extracting this transition can be seen by examining its small angle approximation $\mathcal{O} \simeq \sum_{x=1}^N \theta_x$, which is equal to the phase drop across the small junction, the typical low-energy variable used to describe fluxonium.

The qubit frequency can be extracted via fits of the decomposition of this correlation function as expressed in Eq.~\ref{eq:twopointspec}, and described in detail in Appendix \ref{lat-sim-deets}. For visualisation purposes, it is convenient to construct the \textit{effective frequency}, defined as
\begin{equation}
\label{eq:discrete-eff-freq}
    \nu^{\text{eff}}_{01} = -\frac{1}{2\pi \Delta t} \text{log}\big[ C(t+\Delta t)/C(t) \big]~,
\end{equation}
which converges to the qubit frequency modulo lattice spacing errors as $t\rightarrow \infty$. An alternative is the cosh-corrected effective frequency, defined implicitly as:
\begin{equation}
\label{eq:cosh-corr-discrete-eff-freq}
\frac{C(t)}{C(t + \Delta t)} = \frac{\cosh \left[2\pi \,\nu^\mathrm{eff}_{01} \cdot (t - N_T/2) \right]}{\cosh \left[ 2\pi \, \nu^\mathrm{eff}_{01} \cdot  (t + \Delta t - N_T/2) \right] }~,
\end{equation}
which accounts for the leading thermal behaviour, and converges to the true frequency at a given lattice spacing in the limit that both $t$ and $\beta - t$ approach infinity. 
A plateau in either effective frequency definition will correspond to the qubit frequency of the lattice theory.

\begin{figure}[t!]
    \centering
    \includegraphics[width=0.95\linewidth]{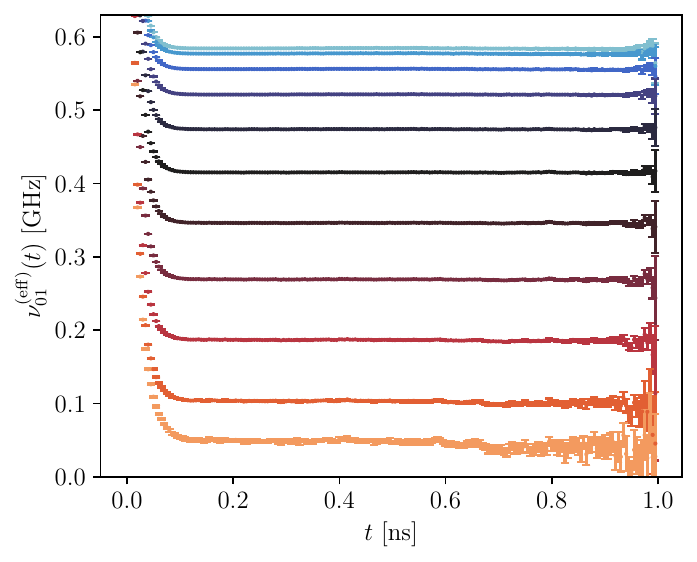} 
    \caption{\textbf{\textit{Effective frequency plot}}: Plot of the cosh-corrected effective frequency, \eqref{eq:cosh-corr-discrete-eff-freq}, for the qubit parameters of \eqref{eq:tn-comp-params} for the interpolating operator $\mathcal{O} = \sum_x \text{sin}\theta_x$ for the ensemble with the coarsest lattice spacing in \Cref{tab:cont-lim-details}. Eleven equally spaced  gate charges are considered, with the blue data corresponding to $n_{g} = 0$ and the orange to $n_{g} = 1/2$, and error-bars have been estimated with $100$ bootstrap resamplings of the Monte Carlo data. }
    \label{fig:correlator}
\end{figure}

An example of the cosh-corrected effective frequency for the qubit defined in \eqref{eq:tn-comp-params} is shown in \figref{fig:correlator}. In this calculation, $n_{gx} = n_g$ is constant along the array, and eleven equally-spaced gate charges are explored. Gate charges are included by reweighting an $n_g=0$ calculation, as described in Section \ref{sec:lattice-formulation}. Clear plateaus are seen, and the qubit frequency of the lattice theory as extracted from the fits described in Appendix \ref{lat-sim-deets} is consistent with the observed plateaus. It is seen that the qubit frequency monotonically decreases as the gate charge increases, which is the same behaviour seen in TN studies~\cite{Di_Paolo_2021}.

\begin{figure}[t!]
    \centering
    \includegraphics[width=0.95\linewidth]{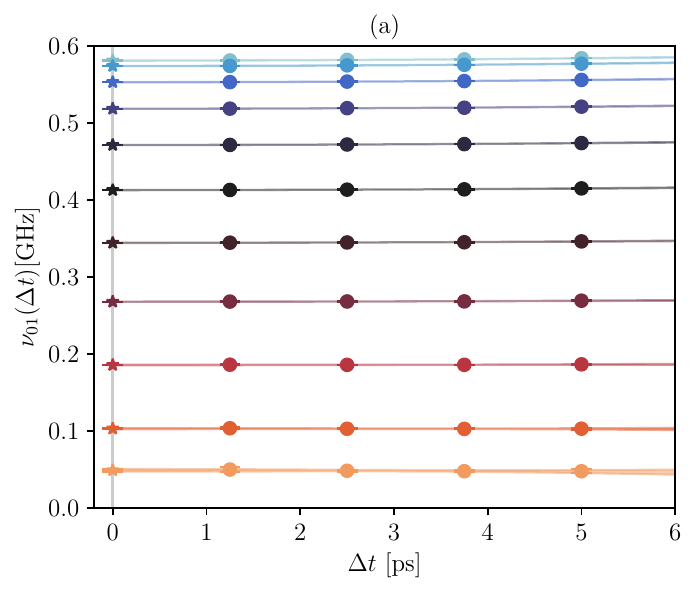}
    \includegraphics[width=0.95\linewidth]{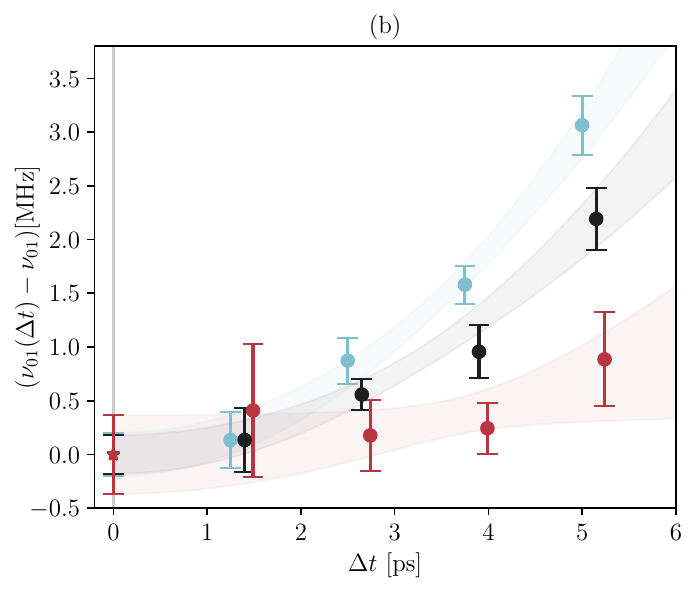}  
    \caption{\textbf{\textit{Continuum limit extrapolation}}: \textit{ (a)} Qubit frequency as a function of lattice spacing for the system defined in \eqref{eq:tn-comp-params}. Computations are performed at 11 equally-spaced points between $n_g=0$ (light blue) and $n_g=1/2$ (light orange points). Error bars are statistical. The flatness of the curves indicate very small lattice spacing artifacts. \textit{(b)}    
    Lattice spacing error of the qubit frequency, i.e. the difference between individual simulations and the continuum limit, for $n_g \in \{0.00,0.25,0.40\}$. 
    The observed quadratic dependence is expected from theory. Color coding is the same as in the top panel. Points are slightly shifted horizontally for visibility.}
    \label{fig:cont-lim}
\end{figure}

Comparison of the lattice results to the TN study requires a continuum limit extrapolation of the lattice results. It can be shown that
\begin{equation}
\label{eq:cont-limt-qubit-freq}
    \nu_{01}(\Delta t) = \nu_{01} + b\Delta t^2 + \mathcal{O}(\Delta t^4)~,
\end{equation}
where $\nu_{01}(\Delta t)$ is the frequency measured on lattice geometries with a finite lattice spacing, and $\nu_{01}$ is the true qubit frequency of the many body theory. Thus, to extract $\nu_{01}$, the effective frequency is extracted for a fixed Hamiltonian on a sequence of lattice geometries with $\Delta t$ ranging over a factor of 4, with the results fit to a quadratic polynomial. The parameters of the ensembles used and details of the ensemble generation are given in \Cref{tab:cont-lim-details}. The dominant error in the continuum-extrapolated lattice calculations is statistical.

The results of the continuum limit are shown in \figref{fig:cont-lim}. The observed lattice spacing errors are on the order of a few MHz for all gate charges, and as expected from theory there is a quadratic dependence on the lattice spacing. The continuum extrapolated qubit frequencies have a precision of $100-500$ kHz. The precision achieved here may be high enough to resolve qubit-array mode couplings (see e.g. \cite{Viola_2015,rui_aps2025}). 

\begin{figure}[t!]
    \centering
    \includegraphics[width=0.95\linewidth]{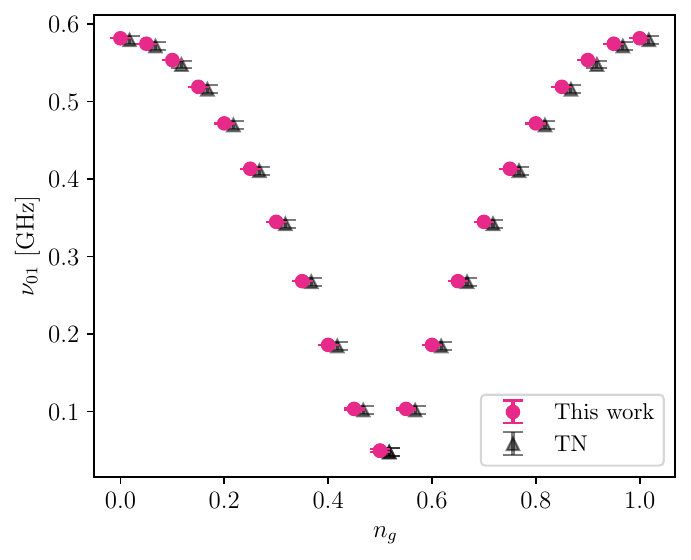}
    \caption{\textbf{\textit{Tensor network comparison}}: Qubit frequency as a function of gate charge $n_g$ for the system defined in \eqref{eq:tn-comp-params}. Continuum-extrapolated lattice results are shown with pink circles while TN results are denoted by gray triangles, and have been slightly shifted in $n_g$ for viewing convenience. Uncertainties on the lattice calculations are statistical while those on TN result from a webplot digitization of the figures reported in Ref.~\cite{Di_Paolo_2021}.}
    \label{fig:ng-sweep}
\end{figure}

In \figref{fig:ng-sweep} the continuum extrapolated qubit frequencies as a function of gate charge are compared with TN results. The qubit frequency is periodic in $n_g$ with period 1 and is symmetric about $n_g=1/2$; calculations are therefore performed up to $n_g=1/2$, with the data reflected about this mid-point. Errors on the TN data shown above come from a Webplot digitization of the results in \cite{Di_Paolo_2021}, and do not reflect the systematic error of the TN results, which are likely much smaller than those shown. Precise numerical agreement is found for the entire range of gate charges, providing the first independent check of Ref.~\cite{Di_Paolo_2021}.

\subsection{Study 2: Charge-noise dephasing}
\label{sec:study2}

This study is designed to parallel the investigations of Ref.~\cite{randeria2024dephasing} where six fluxonium devices with (nearly) constant $E_J,E_C,E_L, \hbar \omega_{pl}$ were studied as a function of the reduced impedance $z$. A primary goal was to study dephasing due to coherent quantum phase slips and the concomitant Aharonhov-Casher effect \cite{PhysRevLett.53.319, PhysRevLett.88.050403}. It was found that the dephasing rate varied by several orders of magnitude as the impedance was varied between $0.057 \leq z \leq 0.101$. The experimental data is consistent with the current theoretical model for the dephasing rate due to coherent quantum phase slips $\Gamma$, which is
\begin{equation}
\label{eq:WKB-spread-Gamma}
    \Gamma = \pi \sqrt{N} \, \epsilon_{ps} |\mathcal{F}_{01}| / h~,
\end{equation}
where $N$ is the number of array junctions,
\begin{equation}
\label{eq:eps_ps}
    \epsilon_{ps}/h = \frac{4 \sqrt{2}}{\pi} \hbar \omega_{\text{pl}} \frac{e^{-\frac{4}{\pi z}}}{\sqrt{z}} / h~,
\end{equation}
is the ``phase slip rate"  of individual array junctions, and the ``structure factor" $ \mathcal{F}_{01} =  \langle 1 | e^{-2\pi i n}| 1 \rangle - \langle 0 | e^{-2\pi i n}| 0 \rangle $ is a difference of $2\pi$ shift matrix elements between the $|0\rangle$ and $|1 \rangle$ state of the superinductance model determined by $(E_J,E_C,E_L, \varphi_{\text{ext}})$. \eqref{eq:WKB-spread-Gamma} results from analyzing an effective Hamiltonian of fluxonium, written in the basis of states labeled by the number of quantum phase slips which have crossed the fluxonium loop \cite{Matveev_2002, manucharyan2012superinductance,PhysRevB.85.024521,randeria2024dephasing}.

A complementary study is performed here using lattice field theory. Calculations are performed of the properties of a sequence of qubits with $E_J,E_C,E_L, \hbar \omega_{pl}$ tuned to approximately the same values as in Ref.~\cite{randeria2024dephasing}, but with higher array junction impedance. The aim is to quantitatively probe phase slips in  a new regime. The parameters of the lattice calculations are fixed at $(E_J,E_C,E_L,\hbar \omega_{pl}) \equiv  (E_J^b, e^2/(2C^b),E^a_J/N,\sqrt{8 E^a_C E^a_J}) = (1.41,3.22,0.25,13.49)$ [h GHz], and computations are performed with $z\equiv \pi^{-1}\sqrt{2 E^a_C / E^a_J}$ values tabulated in \Cref{tab:fluxonium-parameters}. 
The $z=0.101$ qubit simulated here has the same fine-grained circuit parameters as those reported of Q1 in Ref.~\cite{randeria2024dephasing}, providing a point of direct comparison. The fine-grained Hamiltonian parameters taken in the calculations presented are specified in \tableref{tab:fluxonium-parameters}, and the details of the numerical calculations are explained in Appendix~\ref{lat-sim-deets}. In brief, the same correlator method, interpolating operator, and reweighting method of Sec.~\ref{sec:study1} are applied. Results shown in this section are at a fixed $\Delta t = 5$~ps, however a continuum limit was taken for the $z = 0.101$ ensemble by generating an additional ensemble with $\Delta t = 10$~ps, and systematic uncertainties in the charge noise dephasing rate due to non-zero lattice spacings are seen to be around $2\%$.

\begin{table}[b]
\centering
\begin{tabular}{c||c|c|c|c|c} 
 $z$ & $0.101$ & $0.11$ & $0.12$ & $0.13$ & $0.135$  \\
 \hline
$N$ & 85 & 78 & 72 & 66 & 64  \\
$C^a$ [fF] & 18.11	& 16.63 & 15.24 & 14.07 & 13.55  \\
$E_J^a$ [h GHz]  & 21.25 & 19.51 & 17.89 & 16.51 & 15.90  \\
\end{tabular}
\caption{\textbf{\textit{Fluxonium parameters used in Study 2}: }  The first column corresponds to fine-grained device parameters of a system studied in  Ref.~\cite{randeria2024dephasing} while the remaining columns correspond to new parameters. The small junction parameters are held fixed at $C^b = 13.7376 $ [fF] and $E^b_J = 3.22$ [h GHz] for all devices. These choices result in the following qubit parameters $(E_C,E_J,E_L) := (e^2/(2C^b),E^b_J,E^a_J/N) = (1.41,3.22,0.25)$ [h GHz]. The plasma frequency is fixed at $\hbar \omega_{pl} = \sqrt{8 E^a_J E^a_C} = 13.4843$ [h GHz] . In all calculations, ground capacitances are set to zero.
}
\label{tab:fluxonium-parameters}
\end{table}

A current hypothesis in the theoretical model of the CQPS dephasing rate is that it results from phase slips interfering via the Aharonhov-Casher effect due to random, locally varying gate charges, where any configuration of gate charges is considered equally likely. This interference broadens the qubit transition, and the dephasing rate is related to the standard deviation of the qubit frequency distribution $\sigma$ by
\begin{equation}
\label{eq:gamma-sigma}
    \Gamma = \sqrt{2}\pi  \sigma~.
\end{equation}
This scenario is probed here by computing qubit frequency distributions resulting from $1024$  randomly drawn gate charges distributed (spatially) independently and uniformly over $0\leq n_{gx} \leq 1$. The use of reweighting reduces the cost of this exercise by a factor of $\sim1000$ relative to independent qubit calculations, making it computationally feasible. We are not aware of an analogous technique in TN.

\begin{figure}[t!]
    \centering
    \includegraphics[width=0.85\linewidth]{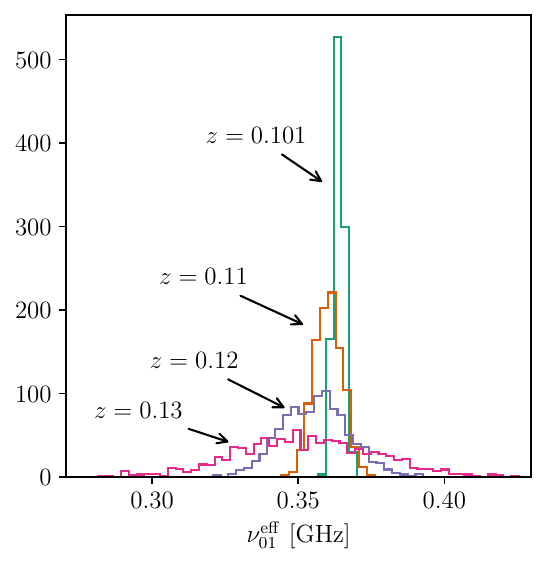} 
    \caption{\textbf{\textit{Qubit frequency distributions}}:  Histogram of qubit frequencies for the device parameters in Table \ref{tab:fluxonium-parameters}. The $z=0.135$ distribution (not shown) has such large spread that it is approximately uniform on the scale of this figure. Histograms are generated from 1024 random draws of the gate charge distribution where all $n_{gx}$ are drawn independently and uniformly over the interval $[0,1)$. Data shown is for fixed $\Delta t = 5$ ps. 
    }
    \label{fig:mallika-comparison1}
\end{figure}

The resulting qubit frequency distributions are shown in \figref{fig:mallika-comparison1}. The increase of the distribution width with impedance is expected from \eqref{eq:eps_ps}, and results in a growing dephasing rate. The decrease in the mean value of the qubit frequency distribution is interesting and perhaps surprising, and shows that even when nominal values of $(E_J, E_C, E_L)$ are held fixed, the static spectral properties of fluxonium are not constant.

Results already present in the literature suggest this observation. A perturbative analysis of many-body interactions in fluxonium predicts that, even at at zero ground capacitance (the scenario explored here), a dominant effect of  ``array mode" interactions is to dress the inductive energy as follows \cite{PhysRevX.3.011003,Viola_2015,Sorokanich:2024wkx,Di_Paolo_2021}:
\begin{equation}
\label{eq:el-renorm}
    E_L=\frac{E_J^a}{N}(1-\frac{\pi z}{2}\frac{N-1}{N}).
\end{equation}
Applying this scheme to single-variable superinductance models produces $z-$dependent qubit frequencies in rough agreement with the lattice data. While a quantitative comparison with experiments would require further knowledge of fine-grained circuit parameters such as ground capacitances, the calculations suggest that $E_L$ can be changed by a few percent at typical fluxonium fabrication parameters.

\begin{figure}[t!]
    \centering
    \includegraphics[width=0.95\linewidth]{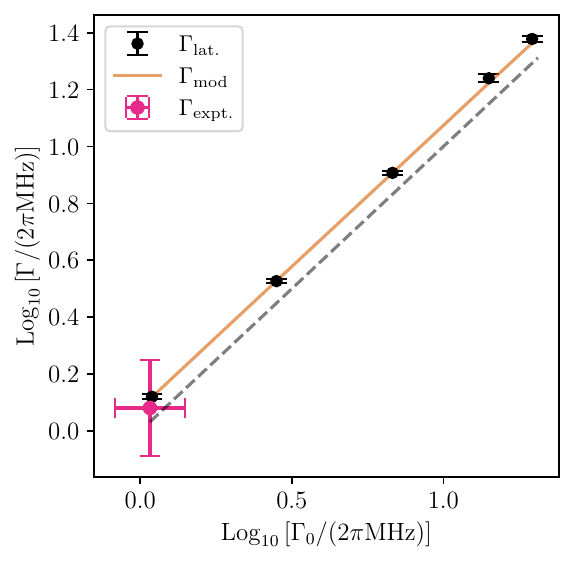} \\
        \caption{\textbf{\textit{Dephasing rates}}: Comparison of dephasing rate computed by different methods. The points from left to right correspond to $z=0.101,0.11,0.12,0.13,0.135$ and the dashed line has unit slope. For each point, the $x$-value is the width computed from \eqref{eq:WKB-spread-sig}, and the $y$-value is the width computed from lattice (black points) or from \eqref{eq:WKB-spread-mod} (orange). The pink point includes values reported for the $z = 0.101$ qubit studied in  \cite{randeria2024dephasing}.  
        }
    \label{fig:mallika-comparison2}
\end{figure}

The current theoretical model for the width of the qubit frequency distribution is \cite{PhysRevB.85.024521}
\begin{equation}
\label{eq:WKB-spread-sig}
    \sigma = \sqrt{N/2} \, \epsilon_{ps} |\mathcal{F}_{01}| / h~.
\end{equation}
The widths in \figref{fig:mallika-comparison1} are close but not equal to these predictions, as shown in \figref{fig:mallika-comparison2}. To explain this, a modified estimate for $\sigma$ is proposed which incorporates several effects, which will be discussed in detail in the following paragraphs. As shown in the orange curve in \figref{fig:mallika-comparison2}, the modified estimate for $\sigma$ agrees quantitatively well with the lattice results for the range of impedances studied. The experimental results of Ref.~\cite{randeria2024dephasing} are consistent with the lattice simulations where a comparison can be made.

The modified prediction for $\sigma$ is detailed below; more details are found in Appendix \ref{new-cqps}. The width of the qubit frequency distribution is estimated by
\begin{equation}
\label{eq:WKB-spread-mod}
    \sigma_{\text{mod}} = \sqrt{N/2} \, \epsilon_{ps}' |\mathcal{F}_{01}'| / h
\end{equation}
where $\mathcal{F}_{01}'$ is the structure factor computed with $E_L$ dressed according to \eqref{eq:el-renorm}, and $\epsilon_{ps}'/h$ is a modified phase slip rate, equal to the frequency  of the $0\rightarrow1$ transition of the model
\begin{equation}
\label{eq:phase-slip-ham}
    H = 4 E_C' n^2- E_J^a \text{cos}(\theta) + \frac{1}{2}E_L'(\theta - \varphi_{ext})^2
\end{equation}
where $E_C'=e^2/(2C_{\text{eff}})$,
\begin{align}
    C_{\text{eff}} & = C^a\big(1 + \frac{N-1}{N^2}\big) + \frac{C^b}{N^2} \nonumber \\
    E_L' & = \frac{E_J^a}{N} \frac{N-1}{N} + \frac{E_J^b}{N^2}.
\end{align}
This energy difference is computed by exact diagonalization. This scheme results in the orange line in \figref{fig:mallika-comparison2}.

This prescription is actually a minimal modification to the usual formula for $\epsilon_{ps}$ \eqref{eq:eps_ps}, which is an analytic approximation of the $0\rightarrow1$ transition frequency of the model
\begin{equation}
\label{eq:phase-slip-ham-manucharyan}
    H = 4 (\frac{e^2}{2 C^a}) n^2- E_J^a \text{cos}(\theta) + \frac{1}{2}(E_J^a/N)(\theta - \varphi_{ext})^2~,
\end{equation}
derived from properties of Mathieu functions. As discussed in Sec. 1.2.3 of Ref.~\cite{manucharyan2012superinductance}, this Hamiltonian models the environment of an array junction undergoing a phase slip event. 
In short, substituting the analytic approximation for a numerical Schrodinger equation solve, incorporating the small junction, and dressing $E_L$, produces agreement with the lattice results.

 \begin{figure}[t!]
    \centering
    \includegraphics[width=1\linewidth]{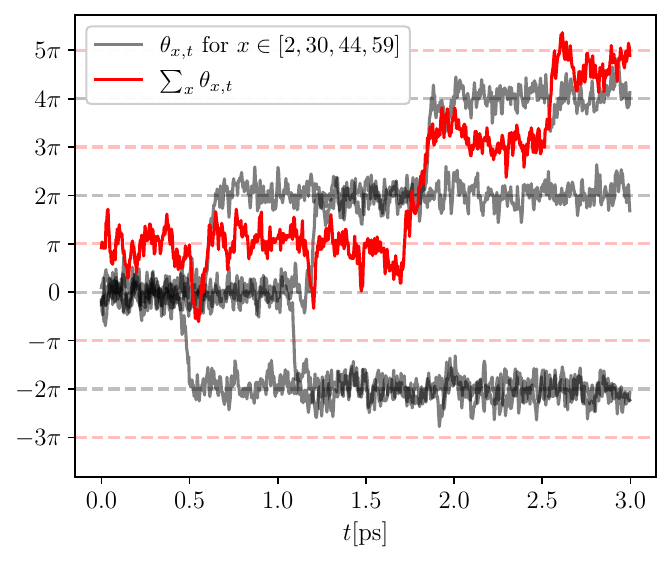}
        \caption{\textbf{\textit{Instantons}}: Euclidean path containing instantons. The horizontal axis is Euclidean time and the vertical axis shows the value of various projections of the  circuit history. The red curve shows the value of $\sum_x \theta_x$ while the black curves show $\theta_2, \theta_{30},\theta_{44},\theta_{59}$.  This example configuration comes from the $z = 0.14$ qubit on a lattice with $\Delta t =5$ ps. Circuit parameters are given in \tableref{tab:fluxonium-parameters} and  for Monte Carlo ensemble data is given in \tableref{tab:cont-lim-details}.
    }
    \label{fig:instanton-statistics}
\end{figure}

A distinctive feature of the lattice approach is its ability to explicitly analyze ``instantons". Exemplified in \figref{fig:instanton-statistics}, instantons are Euclidean-time circuit paths that involve the tunneling between potential energy minima. Instantons carry topological charge and are associated with phase slip events: as shown previously the phase slip amplitude $\epsilon_{ps} \propto e^{-\frac{4}{\pi z}}$ is proportional to the probability of a 1-instanton field configuration, which has Euclidean action $S_E \sim \frac{4}{\pi z}$ \cite{Matveev_2002}. Paths containing more than one tunneling event, often called ``multi-instanton" configurations in high-energy physics, are also possible and examples are seen in \figref{fig:instanton-statistics}. An $n$-instanton event has Euclidean action $S_E \sim n\frac{4}{\pi z}$, and thus at small $z$ are extremely improbable. However as $z$ increases multi-instanton configurations become more probable. The effect of these configurations are neglected in the expression for $\epsilon_{\text{ps}}$.

\color{black}

\color{black}

\subsection{Study 3: Ground capacitance}
\label{sec:study3}

This study examines ground capacitance effects in fluxonium. At non-zero ground capacitance, many couplings between the superinductance mode and array modes become non-zero \cite{symmetries-and-collective,Viola_2015,Sorokanich:2024wkx}. This induces a number of effects on the superiductance mode, including shifts and hybridization of energy levels, which can be detrimental to qubit performance. Here their effect on the qubit frequency and its distribution with gate charge are studied.  

The study consists of a sweep over ground capacitances $0.0 \leq C_g^a = C_g^b\leq 1.0 \ \mathrm{fF}$ at the parameters of the $z=0.101$ qubit investigated in the previous section. Previously, $z$ was varied at fixed (zero) ground capacitance; here $C_g$ is varied, and $z$ is fixed.  The same techniques as described in the previous two sections are used to compute observables. For calculation details see \tableref{tab:fluxonium-parameters} \& Appendix \ref{lat-sim-deets}.

\begin{figure}[t!]
    \centering
    \includegraphics[width=0.9\linewidth]{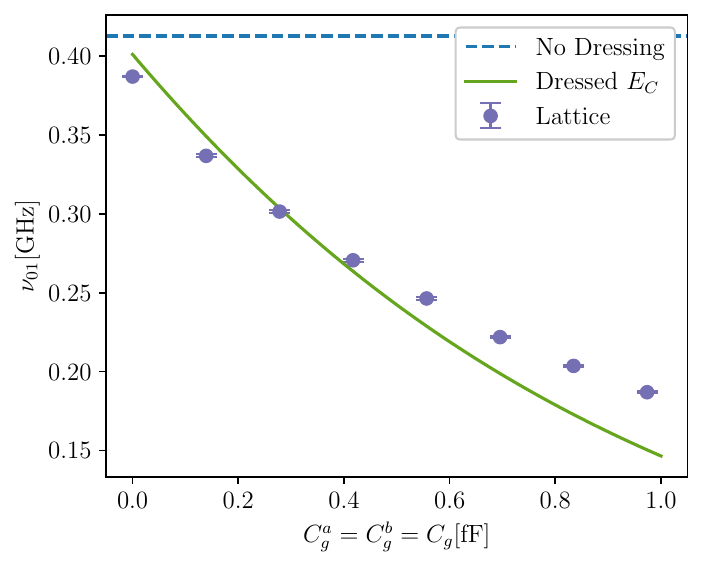}\\
    \caption{\textbf{\textit{Qubit frequency}}: Qubit frequency as a function of ground capacitance, where $C_g^a=C_g^b$ and all gate charges are set to zero. The error bar indicates the statistical error, for fits performed at $\Delta t = 5\mathrm{ps}$. Details of the ensemble are given in Appendix \ref{lat-sim-deets}.}
    \label{fig:ground-cap-qubit-freq}
\end{figure}

\begin{figure}[t!]
    \centering
    \includegraphics[width=0.9\linewidth]{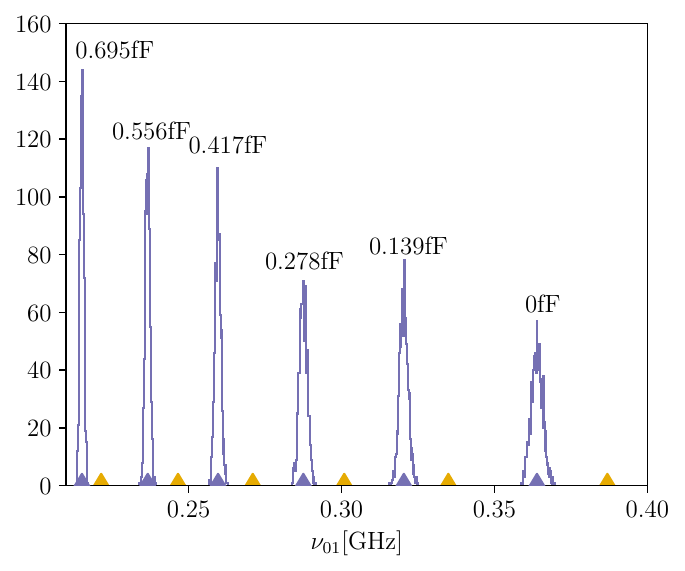}\\
    \caption{\textbf{\textit{Charge dispersion}}: Qubit frequency distributions for six different values of the ground capacitance. Distributions result from 1024 random draws of the gate charge. Purple points indicate the mean of the distribution and yellow is the qubit frequency when all gate charges are zero. Data is shown for fixed $\Delta t = 5$ ps. Details of the ensemble are given in Appendix \ref{lat-sim-deets}. }
    \label{fig:ground-cap-charge-dispersion}
\end{figure}

The results are summarized in \figref{fig:ground-cap-qubit-freq} \& \figref{fig:ground-cap-charge-dispersion}. The qubit frequency and charge dispersion are seen to depend on the ground capacitance. These effects are not included in typical fluxonium models, where neither $(E_J,E_C,E_L) = (E_J^b, e^2/(2C^b),E_J^a/N)$ of the superindcutance mode, nor the equations of the charge dispersion model \eqref{eq:WKB-spread-Gamma} $\&$ \eqref{eq:eps_ps}, depend on ground capacitance. Qualitative features of the trends are attempted to be understood below, however the ground capacitances explored here are likely large enough that present analytic methods fail. This ability to handle arbitrary circuit parameters is a useful feature of the lattice approach.

The qualitative trend in the qubit frequency appears to be explainable with \textit{array mode perturbation theory} \cite{symmetries-and-collective,Viola_2015,Di_Paolo_2021,Sorokanich:2024wkx}. In this framework, the fluxonium Hamiltonian is rewritten in terms of the superinductance mode and $N-1$ \textit{array modes} by choosing a basis of variables which diagonalize the capacitance matrix. By taking into account the full structure of the capacitance matrix, it can be shown that the charging energy of the superinductance mode  is approximately equal to
\begin{equation}
\label{eq:ec-dress-with-cg}
    E_C = \frac{e^2}{2(C^b + \frac{1}{2}C_g^b + C^a/N + (N/12)C^a_g)},
\end{equation}
(see Eq. D6 of \cite{symmetries-and-collective} and \cite{Sorokanich:2024wkx}). This induces an explicit dependence of the qubit frequency on ground capacitance.

As seen in \figref{fig:ground-cap-qubit-freq}, substituting this dressed relation for $E_C$ into the superinductance model accounts for the overall trend of the qubit frequency as a function of ground capacitance. This dressing of $E_C$ is only one effect, and a quantitative comparison requires summing all effects produced by the (many) array mode interactions. While this is beyond the scope of this paper, the lattice approach is thus seen to offer a baseline of comparison to test the efficacy of array mode perturbation theory; no other many-body method has yet offered this option.

The dependence of the charge dispersion on ground capacitance is now explored. \figref{fig:ground-cap-charge-dispersion} shows the qubit frequency distributions for the lowest six values of the ground capacitance, computed from the same Monte Carlo ensembles as used in \figref{fig:ground-cap-qubit-freq}. The width of the qubit frequency distribution decreases as ground capacitances grows. Since the dephasing rate is proportional to this width, this means ground capacitances actually \textit{reduce} charge noise dephasing. This is interesting since ground capacitances are typically seen as detrimental to the qubit. This behavior is not predicted by the standard phase slip formula \eqref{eq:eps_ps}, which is independent of ground capacitance.

There does not appear to be an analytic prediction for the dependence of charge noise on ground capacitance in the literature.  Simply substituting a dressed value of $E_C$ into $\mathcal{F}_{01}$ produces percent-level changes in the width, which does little to explain the factor of $\sim2.5$ decrease. In contrast, a slight generalization of the methods of Ref.~\cite{PhysRevB.102.014512}, which uses periodic gaussian states to analytically estimate charge dispersion, may be able to account for the trend in the charge noise seen in \figref{fig:ground-cap-charge-dispersion}. In the analysis of Ref.~\cite{PhysRevB.102.014512}, ground capacitance effects on the charging energy of array junctions $E_C^a=e^2/(2C^a)$ are ultimately neglected, and the resulting impedance $z=\pi^{-1}\sqrt{2 E_C^a/E_J^a}$ enters the $\epsilon_{ps}$ relation. If at this step ground capacitances were kept, this would lower the charging energy, lowering the impedance and therefore the phase slip rate. This may produce sizable effects since the phase slip rate is proportional to $\text{exp}({-\frac{4}{\pi z}})$. Another interesting feature of the lattice results is that the mean of the qubit frequency distribution is lower than the qubit frequency at zero gate charges. In fact, the effect of gate charges is to \textit{always} lower the qubit frequency.

While adding ground capacitances to TN calculations appears to be straightforward, to date there have been no numerical studies in this direction. This appears to be the first many-body calculation of fluxonium including ground capacitances. Fortunately, in the lattice formalism ground capacitances can be included for essentially no additional computational cost; with or without ground capacitances, the (dense) capacitance matrix appears in the action in the same way. Because of this, it would furthermore be trivial to replace the idealized capacitance matrix model used here (which does not account for disorder in the array junctions, for example) with a realistic capacitance matrix arising from a finite element simulation. The general methods used here may be useful in exploring ground capacitance effects in protected qubits using Josephson junction arrays \cite{PRXQuantum.2.010339,Pechenezhskiy_2020,hays2025non,PRXQuantum.1.010307}.

\section{Discussion}
\label{discussion}

This section presents an analysis of the computational cost and future prospects for the lattice field theory approach to ab-initio calculations of quantum circuits. While the formalism can be used for any lumped element quantum circuit, the studies of fluxonium presented in the previous section provide a numerical baseline that is reasonably general; the number of quantum variables in fluxonium can be varied widely, and fluxonium physics incorporates rich features that require advanced lattice techniques.

The calculations of Sec. \ref{sec:study2}, analyzing systems with $N \sim 50-100$ junctions, required approximately 350 A100 GPU hours to generate the configurations for each ensemble, using an implementation of hybrid Monte Carlo (HMC)~\cite{DUANE1987216} coupled with instanton-inspired updates, as discussed in Appendix \ref{lat-sim-deets}. After the generation of the configurations, subsequent evaluation of physical observables adds negligible computational overhead. This scale of resource usage and throughput is comparable to those reported in state-of-the-art tensor network (TN) studies~\cite{petrescu_2025}.
These calculations are among the largest Monte Carlo–based computations of superconducting circuits reported to date. For comparison, Ref.~\cite{PhysRevB.101.024518} presents simulations of Josephson junction arrays with up to 200 junctions and 3200 temporal slices—the largest currently identified in the literature. Simulations presented here are within a factor of eight of this size. 


The lattice-based framework offers specific advantages over TN methods in selected problem domains. A notable strength is the ability to perform reweighting, which facilitates exploration of nearby configurations in parameter space without requiring new calculations. This capability is particularly advantageous for evaluating the impact of disorder in device architectures, as shown in Sec.~\ref{sec:study2}, and may also support prototyping of circuit designs through reuse of a single reference calculation.
In addition, the approach avoids some limitations of tensor network methods: truncations of local Hilbert spaces may become unjustified in the presence of strong interactions, and the computational cost of TN methods grows significantly when moving beyond one-dimensional geometries.
For other problems, TN approaches are advantageous. In particular, real-time dynamics pose significant challenges for lattice approaches due to the sign problem (although significant efforts in the nuclear and particle physics community seek to change this status~\cite{Gattringer_2016,Wagman:2017gqi, RevModPhys.94.015006,PhysRevD.102.014514,detmold2023signaltonoiseimprovementneuralnetwork,sinclair2018complexlangevinlatticeqcd,PhysRevD.100.114517,larsen2023reducingsignproblemline}).
Specifically, at the current time, the real-time Lindblad evolution analyzed in Ref.~\cite{Di_Paolo_2021} would be computationally prohibitive within the lattice framework presented here. 
At the same time, a similar sampling-based approach to the one developed in this work could be extended to real-time dynamics using variational Monte Carlo (VMC), where approximate wave-function ansätze are sampled while retaining the full local U(1) Hilbert space and avoiding truncations, at the cost of a controlled variational bias, as demonstrated in lattice gauge theory simulations~\cite{periodic_gaussian_2020, lin_2024, bender_2023}.
These contrasts suggest that hybrid approaches, which combine lattice field theory for static observables with VMC or TN methods for dynamic simulations, represent a promising direction. Such integration would allow each method to be applied in contexts where its strengths are most pronounced.

The prospects of calculating the properties of more complex devices with lattice field theory approach are similarly encouraging. A $10 \times 10$ lattice of transmons hosts $\sim100$ quantum variables, which is the roughly the size of present calculations. Thus calculations of the properties of digital quantum computers with $\mathcal{O}(100)$ transmons are of a similar computational scale to those presented here. Lattice methods are routinely applied to compute path integrals in QCD for nuclear and particle physics with as many as $10^9$--$10^{12}$ degrees of freedom; it is to be expected that porting the computational infrastructure developed for this setting to the application to quantum circuits will enable much larger systems to be studied with little additional complexity.

\section{Conclusion}
\label{sec:conclusion}

This work introduces a new approach for computing properties of general superconducting circuits. The method is based on a path integral formulation of circuit-QED, which is then solved with numerical Monte Carlo methods. The presented method for computing Euclidean-time correlation functions enables energies, matrix elements, and other local quantities to be extracted. The most important feature of the method is that it is systematically-improvable and does not require the introduction of systematic errors due to truncation of the infinite-dimensional Hilbert spaces associated with local node fluxes. It is currently the only many-body approach with this feature. Moreover, its efficiency is not restricted to particular circuit geometries, unlike tensor networks, which are primarily suited for one-dimensional systems. The method allows precise calculations of any superconducting circuit formed from a finite number of capacitors, inductors, and Josephson junctions. Demonstration of the method by application to fluxonium reveals several useful features. 

First, the lattice approach can solve many-node circuit QED problems. This is illustrated in Study 1, where previous TN results were reproduced. This feature is particularly useful as circuits grow in size. 

Second, the lattice method can serve to extend approximate theories by offering as a baseline for comparison. This is illustrated in Study 2, where a modified expression for the charge noise dephasing rate is proposed to explain a systematic $20\%$ difference between the lattice results and the current analytical estimate. Within the limited domain explored here, the modified model fits the charge dispersion well. 
To obtain these results an explicit averaging over local charge disorder was performed. This would be difficult to achieve with any other existing method.

Third, the lattice method can be used to probe the effect of many-body interactions on effective circuit parameters. This was illustrated in Study 3, where the effect of ground capacitance on fluxonium was studied. A simple dressed equation for $E_C$ in the superinductance model was seen to explain the qualitative trend of the qubit frequency as a function of ground capacitance. It would be interesting to quantitatively compare the predictions of array mode perturbation theory to those from the lattice approach. It was furthermore found that ground capacitances \textit{lower} the charge noise dephasing rate.

As the fabrication space of fluxonium is further explored, the lattice approach may be applied to other problems. For example, the characterization of fluxonium qubits with very small $E_L$, which requires large inductors where ground capacitance effect accumulate, would benefit from this tool. 
In general, the lattice framework enables many-body computations based on any generic capacitance matrix (e.g., either idealized or from a finite-element simulation) and potential energy of the microscopic circuit. This may be valuable for the development of multi-node circuits utilizing several JJ-array inductors, for example to anticipate parasitic interactions so they can be compensated for before fabrication. Following the general trend of increasing circuit complexity, this approach may be used as a tool to explore new parameter spaces.

Beyond the study of individual qubits, entire processors can be analyzed using the lattice field theory methods proposed here. Crosstalk and other parasitic interactions can be quantified, and signatures of a transition to chaos can be computed within the framework \cite{Berke_2022,blain2025}. Other extended systems, like analog quantum simulators \cite{rosen2024_flatband}, hardware implementations of field theories \cite{ROY2021115445, Pardo_2025}, and quantum amplifiers \cite{9134828, Macklin_science2015}, can also be analyzed.

\color{black}

\acknowledgements 
We thank Steve Girvin, Leonid Glazman, Mallika Randeria, Agustin di Paolo, Thomas Baker, Guifre Vidal, and Alexandru Petrescu for discussions related to this work. 

While at MIT, JL's contributions were supported by the U.S. Department of Energy under Contract No.~DE-SC0011090, by the SciDAC5 award DE-SC0023116, and additionally 
by the National Science Foundation under Cooperative Agreement PHY-2019786 (The
NSF AI Institute for Artificial Intelligence and Fundamental Interactions, http://iaifi.org/). 
While at Argonne National Laboratory, JL's contributions were  
supported by the U.S. Department of Energy, Office of Science, Office of Nuclear Physics through Contract No.~DE-AC02-06CH11357.

M. H. was supported by an appointment to the Intelligence Community Postdoctoral Research Fellowship Program at the Massachusetts Institute of
Technology administered by Oak Ridge Institute for Science and Education (ORISE) through an interagency
agreement between the U.S. Department of Energy and
the Office of the Director of National Intelligence (ODNI).
This research was funded in part by the U.S. Army Research Office under Award No. W911NF-23-1-0045. 

S.S. is supported by the NIST/NRC Postdoctoral Associateship Program.

J.B. acknowledges support by the Alexander von Humboldt Foundation through a Feodor Lynen Research Fellowship.

P.E.S. is supported in part by the U.S.~Department of Energy, Office of Science, Office of Nuclear Physics, under grant Contract Number DE-SC0011090, by Early Career Award DE-SC0021006, by Simons Foundation grant 994314 (Simons Collaboration on Confinement and QCD Strings), by the U.S. Department of Energy SciDAC5 award DE-SC0023116, and has benefited from the QGT Topical Collaboration DE-SC0023646. P.E.S. thanks the Institute for Nuclear Theory at the University of Washington for its kind hospitality and stimulating research environment. This research was supported in part by the INT's U.S. Department of Energy grant No. DE-FG02-00ER41132.

N.C.W. is supported by: the U.S. Department of Energy, Office of Science under grant Contract Numbers DE-SC0011090 and DE-SC0021006, the Simons Foundation grant 994314 (Simons Collaboration on Confinement and QCD Strings), and the U.S. Department of Energy, Office of Science, National Quantum Information Science Research Centers, Co-Design Center for Quantum Advantage under Contract No. DE-SC0012704.

\appendix

\section{Lattice calculation details}
\label{lat-sim-deets}

For each of the studies presented in the main text, the  Hybrid-Monte-Carlo (HMC)  algorithm \cite{DUANE1987216} was used to sample from the quantum mechanical path integral \eqref{eq:fund-expression}. 
For a given action $S(\theta)$ where $\theta \in (S^1)^N$ is a vector of $N$ angles, the HMC algorithm introduces conjugate momenta $p \in \mathbb{R}^N$ with an associated Hamiltonian:
\begin{equation}
H(p,\theta) = \frac{1}{2} p^T p + S(\theta), 
\end{equation}
and equations of motion:
\begin{equation}\label{eq:eom}
\frac{\mathrm{d} \theta}{\mathrm{d} t_{\mathrm{MD}}} = \frac{\partial H }{\partial p} = p, \quad \frac{\mathrm{d} p}{\mathrm{d} t_{\mathrm{MD}}} = - \frac{\partial H}{\partial \theta} = -\frac{\partial S}{\partial \theta}, 
\end{equation}
where the subscript MD is used to distinguish the molecular-dynamics time from Euclidean time. 
To generate the next sample $\theta_{i+1}$ in the Markov chain from the previous sample $\theta_i$, momenta are drawn from the gaussian distribution $\text{prob}(p) \propto\mathrm{exp} \left[ - \frac{1}{2} p^T p \right]$, then the equations of motion shown in \Cref{eq:eom} are used to evolve $\{p,\theta_i\}$ for a fixed molecular dynamics time $t_\mathrm{MD}$. 
The final momenta $p(t_\mathrm{MD})$ are discarded and $\theta_{i+1} = \theta(t_\mathrm{MD})$. 
Iterating this procedure generates a Markov chain that is distributed according to the desired distribution $p(\theta) \sim \frac{1}{Z} e^{-S(\theta)}$. 

In practice, the \Cref{eq:eom} is discretised with the second-order leapfrog (or St\"ormer-Verlet) integrator with stepsize $\epsilon$ for a total of $N_\mathrm{LF} = \frac{t_\mathrm{MD}}{\epsilon}$ steps \cite{PhysRev.159.98,Hairer_Lubich_Wanner_2003}.
Due to numerical integration errors, the Hamiltonian is not exactly preserved by leapfrog integration, and the resulting configurations are only accepted if a uniformly generated number between $0$ and $1$ is less than $\mathrm{exp}\left[-\left(H(p_{i+1},\theta_{i+1}) - H(p_i,\theta_i) \right)\right]$, otherwise $\theta_{i+1}$ is set to $\theta_i$ and the update is rejected. 

As the continuum limit is approached, the path integrals described in this paper concentrate into different topological sectors, leading to diverging autocorrelation times in the Monte Carlo chain. 
These sectors correspond to the global winding number for each angle $N_x$, as well as the local winding number corresponding to the collective mode $\sum_{x=1}^N \theta_x$:
\begin{equation}\label{eq:winding}
N_x = \frac{1}{2 \pi}\int_0^{1/T} \frac{\mathrm{d} \theta_x}{\mathrm{d} t}\ \mathrm{d} t , \quad N_c = \frac{1}{2 \pi}\int_{t_0}^{t_1} \frac{\mathrm{d} \sum_{x=1}^N \theta_x}{\mathrm{d} t} \ \mathrm{d} t.
\end{equation}
Note that as $N$ (the number of array junctions) becomes large, it is possible over a Euclidean time range $[t_0,t_1]$ for all angles to be approximately constant, and still for the collective mode $\sum_{i=1}^N \theta_i$ to increase by $2 \pi$, hence leading to the distinction between the global and local winding numbers shown in \Cref{eq:winding}. 

To mitigate the large autocorrelation times due to needing to thermalise between the different topological sectors, manual instanton updates are proposed. 
The first type depends on hyperparameters $r,w$, parametrising the steepness of the shape of the instanton and the width of its insertion into the configuration. 
The proposal first generates a random insertion time $1 \leq t_0 \leq N_t$, a random angle index $1 \leq i_0 \leq N$, and a random sign $s \in \{+1,-1\}$. 
Then for $\theta_{i_0}(t)$ in the window $t_0 \leq t < t_0 + w$ (where it is understood that if $t_0 + w > N_t$, the instanton wraps around the periodic time dimension) there is a Metropolis-Hastings proposal given by:
\begin{equation}\label{eq:i1}
\theta_{i_0}(t) \mapsto 
    \theta_{i_0}(t) + 2 \pi s\ \frac{\tanh(r(t-t_0)) + 1}{2} .
\end{equation}
The second type of proposal corresponds to collective mode instantons, and also requires a random insertion time $1 \leq t_0 \leq N_t$. 
The hyperparameters are the same $r$ hyperparameter as in \Cref{eq:i1}, and two positive integer widths $w_1,w_2$.  The proposal is given by changing \textit{all} angles $\theta_i(t)$ in the windows $W_1 = [t_0,t_0+w_1), W_2 = [t_0+w_1,t_0+w_1+w_2), W_3 = [t_0+w_1+w_2,t_0+2w_1+w_2)$ by:
\begin{equation}
\theta_{i}(t) \mapsto
\begin{cases}
    \theta_{i}(t) + \frac{2 \pi s}{N}\ \frac{\tanh(r(t-t_0)) + 1}{2}  & t \in W_1 \\
    \theta_{i}(t) + \frac{2 \pi s}{N}  & t \in W_2 \\
    \theta_{i}(t) + \frac{2 \pi s}{N}\ \frac{\tanh(-r(t-t_0-w_1-w_2)) + 1}{2}  & t \in W_3 \\
\end{cases}
\end{equation}
corresponding to the insertion of a collective instanton anti-instanton configuration. 

After block-averaging the correlation functions to mitigate the autocorrelation times, the sample mean and covariance was determined by averaging over the $N_\mathrm{block}$ resulting measurements:
\begin{align}
\hspace{-1cm}\overline{C(t)} &= \frac{1}{N_\mathrm{block}}\sum_{i=1}^{N_\mathrm{block}} C_i(t) \nonumber \\ \qquad \hspace{-0.8cm}S(t_1,t_2) &= \sum_{i = 1}^{N_\mathrm{block}} \left(C_i(t_1) - \overline{C(t_2)} \right) \left( C_i(t_2) - \overline{C(t_2)}\right) ~.
\end{align}
Ledoit-Wolf optimal shrinkage \cite{LEDOIT2004365} 
was applied to the extracted covariance matrix due to the large dimensionality of the space. 
In principle, fitting the spectral decomposition of \Cref{eq:twopointspec} requires a choice of fitting range, as well as a number of states to include in the fit. 
Ideally, fitting ranges are chosen in regions where the effective mass shows a plateau and the correlation function can reasonably be fit with one or two states.
Note that in all three studies, the operator used $\sum_{x=1}^N \sin(\theta_x)$ is charged under the parity symmetry $P : \vec{\theta} \to -\vec{\theta}$, thus the operator only couples $P$-even states to $P$-odd states. 
Thus, the spectral decomposition can be written as:
\begin{align}
&C_{{e},{o}}(t) = \frac{1}{\sum_{n=1}^{{e}} e^{-E_n \beta} +  \sum_{m=1}^{{o}} e^{-E_m \beta}} \times \nonumber \\
&\sum_{n=1}^{{e}} \sum_{m=1}^{{o}} |\langle n | \mathcal{O} |m\rangle|^2 \times \nonumber \\
& \Big( e^{-E_n t/\hbar} e^{-E_m(\beta - t/\hbar)}  + e^{-E_n (\beta-t/\hbar)} e^{-E_m t/\hbar}\Big)
\end{align}
where $e,o$ are the number of states used in the $P$-even and $P$-odd sectors. 
For a particular fitting range, and choice of $N_\mathrm{e}, N_\mathrm{o}$, the nonlinear least squares fit minimizes the Pearson-$\chi^2$ of the fit:
\begin{equation}
\chi^2 := \left[\overline{C(t_1)} - C_{e,o}(t_1)  \right] \tilde{S}^{-1} (t_1,t_2)
\left[\overline{C(t_2)} - C_{e,o}(t_2)  \right]
\end{equation}
where sums over $t_1,t_2$ are implicit.
Statistical uncertainties on such a fit are evaluated by bootstrap resampling the mean $\overline{C(t)}$, and performing the fit with the same range, number of states, and shrunk covariance matrix on the bootstrap means.
Statistical errors on any function $f(\mathbf{a})$ of the model parameters $\mathbf{a}$ are determined by:
\begin{align}
&\sigma^2_{f(\mathbf{a}_{\mu})} \nonumber \\
&\quad = \frac{1}{2} \left[Q_{5/6} (f(\mathbf{a}_{b,\mu}) - f(\mathbf{a}_\mu) ) - Q_{1/6} (f(\mathbf{a}_{b,\mu}) - f(\mathbf{a}_\mu) ) \right]
\end{align}
similarly to Ref~\cite{NPLQCD:2020ozd}, where $Q_{\alpha}(x)$ denotes the $\alpha$-quartile values of samples of the random variable $x$, $\mu$ indexes a particular fitting range and choice of $e,o$, and $b$ is a bootstrap index. 
To deal with the model uncertainties arising with the choice of fitting range and number of states used, these hyperparameters were varied over a wide range in each of the studies before being combined with model averaging to provide a determination of the systematic uncertainty. 
Following Ref~\cite{Neil:2022joj}, bayesian model averaging to determine a scalar function  $f(\mathbf{a})$ evaluated on model parameters $\mathbf{a}$ can be written as:
\begin{align}
\langle f(\mathbf{a}) \rangle = \sum_\mu f(\mathbf{a}_\mu^*) \mathrm{Pr}(M_\mu | \{C(t)\}) 
\end{align}
where $M_\mu$ is a list of different models used to fit the correlation function data $C(t)$. 
The resulting uncertainty on the determination is given by:
\begin{align}\label{eq:statsys}
&\sigma^2_{f(\mathbf{a})} = \sum_\mu \sigma^2_{f(\mathbf{a}_\mu)} \mathrm{Pr}(M_\mu | \{C(t)\}) + \nonumber \\
&\sum_\mu f(\mathbf{a}_\mu^*)^2 \mathrm{Pr}(M_\mu | \{C(t)\})  - \left(\sum_\mu f(\mathbf{a}_\mu^*) \mathrm{Pr}(M_\mu | \{C(t)\}) \right)^2
\end{align}
where the first line of \Cref{eq:statsys} is associated with the statistical error, and the second line of \Cref{eq:statsys} is associated with the systematic uncertainty. 
The model weights are chosen according to the Bayesian Akaike Information Criterion \cite{Neil:2022joj} with a uniform prior on the selection of models:
\begin{equation}
\mathrm{Pr}(M_\mu | \{C(t)\}) = \mathrm{exp} \left[-\left( \chi^2 (\mathbf{a}^*) + 2k + 2d_C \right)\right] 
\end{equation}
where $k$ is the number of parameters used and $d_C$ is the number of points measured that have been discarded to form the fitting range chosen. 
This procedure causes models with lower $\chi^2$, lower number of fit parameters and lower number of points discarded to be favoured in the model averaging.

\renewcommand{\arraystretch}{1.5}
\begin{table}[t]
\centering
\begin{tabular}{c|c|c|c|c} 
 $\Delta t$ [ps]  & $N_\mathrm{meas}$ & $\epsilon$ & $w$ & $(w_1,w_2)$\\
 \hline
 5    & $2.3 \cdot 10^{5}$ & 0.025  & 20 & $(5,1),(10,1),(10,2),(10,3),(20,1)$\\
 3.75 & $1.8 \cdot 10^{5}$ & 0.0215 & 30 & $(7,1),(15,1),(15,5),(30,1)$\\
 2.5  & $2.1 \cdot 10^{5}$ & 0.018  & 40 & $(10,1),(20,1),(20,5),(40,1)$ \\
 1.25 & $1.0 \cdot 10^{5}$ & 0.015  & 80 & $(20,1),(40,1),(40,10),(80,1)$ 
\end{tabular}
\caption{\textbf{\textit{Ensembles used in Study 1}: }  For each of these ensembles, a fixed number of time-slices $N_t = 400$ was used, such that the temperature $T = \frac{\hbar}{k_B N_t \Delta t}$ varies by ensemble, where $k_B$ is the Boltzmann constant. A total of $N_\mathrm{meas}$ measurements were taken on each ensemble, which were then block averaged as described in the main text. The number of leapfrog steps used $N_{\mathrm{LF}} = 10$ was held constant, and the stepsize $\epsilon$ was tuned to achieve reasonable acceptance rates. The shape hyperparameter used in both types of instanton updates was also held constant at $r = 2$.
}
\label{tab:cont-lim-details}
\end{table}

\textbf{\textit{Study 1}}: Four ensembles were created at the parameters listed in \Cref{tab:cont-lim-details} for the purposes of taking a continuum limit. 
Instanton updates were proposed every HMC step according to the width parameters $w$ and $(w_1,w_2)$ described in \Cref{tab:cont-lim-details}.
The correlation function $\langle \mathcal{O}(0) \mathcal{O}(t) \rangle$ was measured every $10^3$ HMC steps on 512 parallel HMC chains, where $\mathcal{O}(t) = \sum_{x=1}^N \sin (\theta_x(t))$.
The correlation functions were reweighted according to the procedure described in \Cref{eq:reweighting}, before being block-averaged every $10^3$ measurements to mitigate the autocorrelation times within the Monte-Carlo chains.

\renewcommand{\arraystretch}{1.5}
\begin{table}[t]
\centering
\begin{tabular}{c|c|c|c|c|c} 
$z$ & 0.101 & 0.11 & 0.12 & 0.13 & 0.135   \\ \hline
$N_\mathrm{meas}$ & $9.0 \cdot 10^6$	& $3.6 \cdot 10^6$ & $1.4 \cdot 10^7$ & $1.2 \cdot 10^7$ & $6.6 \cdot 10^7$  \\
\end{tabular}
\caption{\textbf{\textit{Ensembles used in Study 2}: } Due to the tunings of these parameters to have approximately the same effective Hamiltonian for the collective mode, many of the Monte-Carlo parameters can be held fixed. For each of these ensembles, the stepsize $\epsilon = 0.01$, the number of leapfrog steps $N_\mathrm{LF} = 100$, the number of timeslices $N_t = 800$, and the number of HMC updates between measurements (100) was held fixed. Furthermore, instanton injections corresponding to $w = 20$ and $(w_1,w_2) = (15,1)$ were used for all ensembles. More configurations were generated at larger values of $z$ due to larger statistical fluctuations. 
}
\label{tab:fluxonium-parameters2}
\end{table}

\textbf{\textit{Study 2}}: 
The results shown for this study were obtained from ensembles generated for each of the $z$-values listed in \Cref{tab:fluxonium-parameters}, each with a fixed lattice spacing $\Delta t = 5$ps. The details of the Monte-Carlo sampling are described in \Cref{tab:fluxonium-parameters2}. 
A continuum limit is not taken for the dephasing results presented in \Cref{fig:mallika-comparison1,fig:mallika-comparison2}, as additional ensembles with $\Delta t = 10$~ps generated for $z = 0.101, z = 0.135$ found percent-level systematic effects due to the finite lattice spacing, which are ignored. 

\textbf{\textit{Study 3}}: For this study, eight different ensembles
were generated for the $z = 0.101$ parameters described in \Cref{tab:fluxonium-parameters}, corresponding to $C^a_g = C^b_g = n\cdot 0.139 ~\text{[fF]}$ for $n\in\{0,\dots,7\}$. All ensembles have $\Delta t = 5\mathrm{ps}$, $N_t = 800$,  $N_\mathrm{LF} = 100$, $\epsilon = 0.01$. Observables were computed at $\Delta t = 2.5$ ps with lower statistics and did not change at the level of the statistical error.

\section{Quantum phase slip rate model}
\label{new-cqps}

To explain the modifications of the coherent quantum phase slip rate which account for the discrepancy observed in the lattice simulations, it is helpful to briefly review the origin of the array junction phase slip rate $\epsilon_{ps}$, reproduced here for convenience
\begin{equation}
\label{eq:eps-ps-appendix}
    \epsilon_{ps}/h = \frac{4 \sqrt{2}}{\pi} \hbar \omega_{\text{pl}} \frac{e^{-\frac{4}{\pi z}}}{\sqrt{z}} / h~.
\end{equation}
As explained in Sec. 1.2.3 of \cite{manucharyan2012superinductance}, the standard expression for $\epsilon_{ps}$ is derived from an effective hamiltonian
\begin{equation}
\label{eq:phase-slip-ham}
    H = 4 E_C^a n^2- E_J^a \text{cos}(\theta) + \frac{1}{2}\frac{E_J^a}{N-1}(\theta - \varphi_{ext})^2~,
\end{equation}
which models the potential experienced by a specific array junction phase difference $\theta$ which undergoes a quantum phase slip (here $E_C^a=e^2/(2C^a)$). The model  encodes the following reasoning: if it is known that a single phase slip occurs at $\theta$, then this particular quantum variable explores its entire cosine potential while all others experience small fluctuations. Using the path integral formalism, these two assumptions can be used to deduce the effective Hamiltonian above, and the phase slip rate is defined as the $0 \rightarrow 1$ transition rate of this theory. Neglecting the quadratic potential, an analytic approximation of the transition rate can be obtained from the asymptotic properties of Mathieu characteristics, and this analytic estimate is the right hand side of \eqref{eq:eps-ps-appendix}. It is important to note that this phase slip rate assumes an array of identical Josephson junctions.

A more realistic estimate can be obtained by numerically-exactly computing $(E_1 - E_0)/h$ of \eqref{eq:phase-slip-ham}. This modification accounts for some of the discrepancy between the lattice simulation and the analytic prediction. The following model, along with the dressing of $E_L$, produces good agreement.

The model used to predict a modified phase slip rate is derived from examining the Euclidean action of a fluxonium device. For a floating device in the absence of ground capacitances, it reads
\begin{align}
    S_E & = \int dt \Bigg[ \frac{\varphi_0^2}{2} C^a\sum_{x=1}^N  \dot{\theta}_x^2 + \frac{\varphi_0^2}{2} C^b \dot{\theta}_0^2 \nonumber \\
    & + E_J^a \sum_{x=1}^N(1-\text{cos}\theta_x) + E_J^b (1-\text{cos}\theta_0) \Bigg]
\end{align}
where $\theta_0 + \sum_{x=1}^N \theta_x = \varphi_{\text{ext}}$ . Consider a Euclidean time history of the circuit where one and only one of the array junction variables experiences a phase slip. The action has an $S_N$ symmetry corresponding to permutations of the array variables \cite{symmetries-and-collective}, so one is free to choose $\theta_1$ as the junction which slips without changing the result. Then $\theta_1$ experiences its full cosine potential while the other variables fluctuate with small amplitude, even the small junction variable $\theta_0$. Typically the small junction fluctuates the most, but not for the considered path. 

Assuming that all other variables have equal magnitude (which is not in general true), then the fluxoid quantization condition requires that $\theta_0, \theta_{j} = (-\theta_1+\varphi_{\text{ext}})/N$ for all $j\neq i$. Substituting this into the Euclidean action and truncating the small angle approximation of cosine at second order, one obtains the following Euclidean action for a single variable:
\begin{align}
    S_E & = \int dt \Bigg[ \frac{\varphi_0^2}{2} C_{\text{eff}}  \dot{\theta}_1^2+ E_J^a (1-\text{cos}\theta_1) + \frac{1}{2}E_L' (\theta_1 - \varphi_{ext})^2 \Bigg]
\end{align}
where
\begin{align}
    C_{\text{eff}} & = C^a\big(1 + \frac{N-1}{N^2}\big) + \frac{C^b}{N^2} \nonumber \\
    E_L' & = \frac{E_J^a}{N} \frac{N-1}{N} + \frac{E_J^b}{N^2}.
\end{align}
This Euclidean action arises from forming a path integral representation of a single variable system with Hamiltonian 
\begin{equation}
    H = 4 E_C' n_1^2- E_J^a \text{cos}(\theta_1) + \frac{1}{2}E_L'(\theta_1 - \varphi_{ext})^2
\end{equation}
where $E_C'=e^2/(2C_{\text{eff}})$. The array junction phase slip rate is taken to be $(E_1- E_0)/h$ of this model.

\newpage

\bibliography{bib}
\end{document}